\documentclass[iop]{emulateapj}

\usepackage{apjfonts}

\catcode`\@=11
\newcommand{\gapprox}{\mathrel{\mathpalette\@versim>}}
\newcommand{\lapprox}{\mathrel{\mathpalette\@versim<}}
\newcommand{\propapprox}{\mathrel{\mathpalette\@versim\propto}}
\newcommand{\@versim}[2]
  {\lower3.1truept\vbox{\baselineskip0pt\lineskip0.5truept
\ialign{$\m@th#1\hfil##\hfil$\crcr#2\crcr\sim\crcr}}}
\catcode`\@=12

\shorttitle{SUPERNOVA EJECTA IN YOUNGEST GALACTIC SNR G1.9+0.3}

\begin{document}

\title{Supernova Ejecta in the  
Youngest Galactic Supernova Remnant G1.9+0.3}

\author{Kazimierz J. Borkowski,\altaffilmark{1}
Stephen P. Reynolds,\altaffilmark{1}
Una Hwang,\altaffilmark{2}
David A. Green,\altaffilmark{3}
Robert Petre, \altaffilmark{2}
Kalyani Krishnamurthy, \altaffilmark{4}
\&\ Rebecca Willett \altaffilmark{4}
}

\altaffiltext{1}{Department of Physics, North Carolina State University, 
Raleigh, NC 27695-8202; kborkow@unity.ncsu.edu} 
\altaffiltext{2}{NASA/GSFC, Code 660, Greenbelt, MD 20771}
\altaffiltext{3} {Cavendish Laboratory; 19 J.J. Thomson Ave., 
Cambridge CB3 0HE, UK}
\altaffiltext{4}{Department of Electrical and Computer Engineering, 
Duke University, Durham, NC 27708}

\journalinfo{The Astrophysical Journal Letters}
%\submitted{Submitted May 4, 2013}
\submitted{Submitted to ApJ Letters}

\begin{abstract}
G1.9+0.3 is the youngest known Galactic supernova remnant (SNR), with
an estimated supernova (SN) explosion date of $\sim 1900$, and most
likely located near the Galactic Center. Only the outermost ejecta
layers with free-expansion velocities $\ga 18,000$ km s$^{-1}$ have
been shocked so far in this dynamically young, likely Type Ia SNR. A
long (980 ks) {\sl Chandra} observation in 2011 allowed
spatially-resolved spectroscopy of heavy-element ejecta. We denoised
{\sl Chandra} data with the spatio-spectral method of Krishnamurthy et
al., and used a wavelet-based technique to spatially localize thermal
emission produced by intermediate-mass elements (IMEs: Si and S) and
iron. The spatial distribution of both IMEs and Fe is extremely
asymmetric, with the strongest ejecta emission in the northern rim. Fe
K$\alpha$ emission is particularly prominent there, and fits with
thermal models indicate strongly oversolar Fe abundances. In a
localized, outlying region in the northern rim, IMEs are less abundant
than Fe, indicating that undiluted Fe-group elements (including
$^{56}$Ni) with velocities $> 18,000$ km s$^{-1}$ were ejected by this
SN.  But in the inner west rim, we find Si- and S-rich ejecta without
any traces of Fe, so high-velocity products of O-burning were also
ejected. G1.9+0.3 appears similar to energetic Type Ia SNe such as SN
2010jn where iron-group elements at such high free-expansion
velocities have been recently detected. The pronounced asymmetry in
the ejecta distribution and abundance inhomogeneities are best
explained by a strongly asymmetric SN explosion, similar to 
those produced in some recent 3D delayed-detonation Type Ia models.
\end{abstract}

\keywords{
ISM: individual objects (G1.9+0.3) ---
ISM: supernova remnants ---
nuclear reactions, nucleosynthesis, abundances --- 
X-rays: ISM 
}

\section{Introduction}
\label{intro}

\object{G1.9+0.3} is the youngest known Galactic supernova remnant (SNR)
\citep{reynolds08b,green08}, with an estimated supernova (SN)
explosion date of $\sim 1900$, most likely located near the Galactic
center at a distance of 8.5 kpc.  Its spectrum is dominated by
nonthermal continuum well described by synchrotron emission from a
power-law electron distribution with an exponential cutoff
\citep[Paper I;][]{reynolds09}.  Our 237-ks {\sl Chandra} observation
in July 2009 showed conclusively 
thermal emission from ejecta from
the north rim (near the radio, but not the X-ray, maximum), with
strong line emission from intermediate-mass elements (IMEs) such as Si
and S, as well as from Fe \citep{borkowski10}.  In addition, we found
evidence for emission from $^{44}$Sc, produced by the radioactive
decay of $^{44}$Ti.

Only the outermost ejecta layers with free-expansion velocities in
excess of 18,000 km s$^{-1}$ have been shocked so far in this very
dynamically-young SNR \citep{carlton11}.  The presence of Fe at such
high velocities is surprising, since few SNe show such high-velocity
Fe (or $^{56}$Ni). 
The spatial
distribution, abundances, and physical properties of ejecta are key
for understanding the explosion mechanism in this likely Type Ia
SNR. {\sl Chandra} observed \object{G1.9+0.3} for nearly 1 Ms in 2011, which
clarified the $^{44}$Sc detection 
\citep[spectrum of the interior is shown in][]{borkowski13} and provided
detailed images and spectra. Here we report spatially-resolved
spectroscopy of SN ejecta, and interpret the results in the framework
of an energetic and asymmetric Type Ia explosion.

\section{Observations}
\label{obssec}

{\sl Chandra} observed \object{G1.9+0.3} in 2011 May and July (see Table
\ref{observationlog}). The ACIS S3 chip was used in  
Very Faint mode.  Data reduction and spectral extraction were done with
CIAO v4.5 and CALDBv4.5.1.
No significant particle flares were found.  The total duration was 977 ks.

\begin{deluxetable}{lccc}
\tablecolumns{4}
\tablewidth{0pc}
\tablecaption{{\sl Chandra} Observations of G1.9+0.3 in 2011}
\tablehead{
\colhead{Date} & Observation ID & Roll Angle & Effective exposure time \\
%\colhead{Date} & Observation ID & Roll Angle & Effective exposure \\
& & (deg) & (ks) }
%& & (deg) & time (ks) }

\startdata
09--11 May  & 12691 & 79  & 184 \\
12--14 May  & 12692 & 79  & 162 \\
16--17 May  & 12690 & 79  & 48  \\
18--19 May  & 12693 & 79  & 127 \\
20--22 May  & 12694 & 79  & 158 \\
23 May      & 12695 & 79  & 39  \\
14--16 July & 12689 & 277 & 156 \\
18 July     & 13407 & 277 & 48  \\
22--23 July & 13509 & 277 & 55  
\enddata
\label{observationlog}
\end{deluxetable}

In order to examine spatial structures in spectral lines and continua,
data cubes, $256^3$ in size, were extracted from merged event
files. The spatial scale is one ACIS pixel ($0.492\arcsec \times
0.492\arcsec$), while spectral channels from 84 to 595 
were binned by a factor of 2
(these correspond to the energy range of 1.2--8.7
keV, including almost all source photons). 
At the ACIS S3 spectral resolution, the binning introduces negligible
spectral degradation above 1.2 keV.

X-ray spectra from various locations within the remnant were then extracted
from each dataset listed in Table~\ref{observationlog}, and then
summed. The spectral responses were averaged using effective exposure
times as weights. 
The background was modeled rather than 
subtracted, by fitting the spatially-integrated 
spectrum of \object{G1.9+0.3} together with a background spectrum. 
The X-ray analysis was performed with XSPEC v12.7.1
\citep{arnaud96}, using C-statistics (unless mentioned otherwise) 
and APEC NEI v2.0 atomic 
data augmented with
inner-shell processes \citep{badenes06}.  
We also updated inner-shell atomic data for all ions of Si, S, Ar,
and Ca with more than 3 electrons present. Relevant collisional
excitation cross sections and inner-shell ionization cross sections by
electrons were calculated with the Los Alamos Atomic Physics
%Codes\footnote{Access to these codes is provided online at
Codes\footnote{Access to these codes is provided at
aphysics2.lanl.gov/tempweb/lanl/} \citep{mann83,clark91}. Line
energies and branching ratios for radiative transitions are from
\citet{palmeri08}. Only strong resonance lines have been included.

\section{Line Emission}

\begin{figure}
%\epsscale{1.0}
\epsscale{1.2}
\vspace{0.1truein}
\plotone{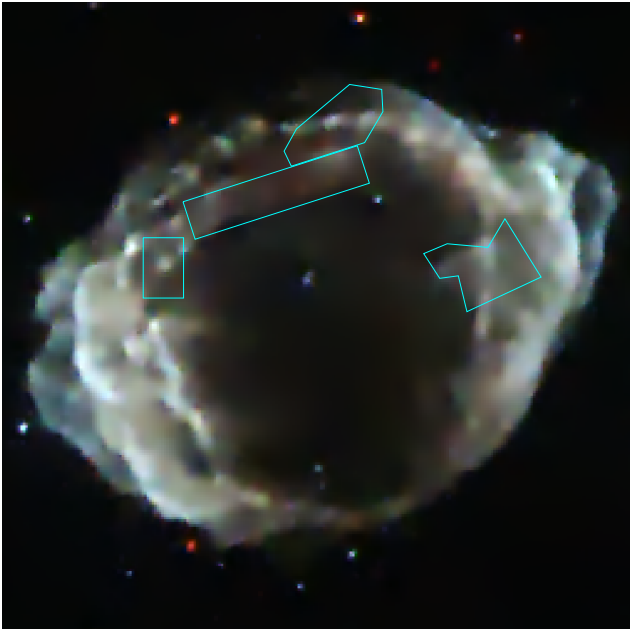}
%\plotone{g1p9threecoloryr11.ps}
\caption{
{\sl Chandra} image of G1.9+0.3, with regions containing spectral lines overlaid. 
Red,
1 -- 3 keV; green, 3 -- 4.5 keV; blue, 4.5 -- 7.5 keV.
Image size $125'' \times 125''$. 
N is up and E is to the left.
\label{2xim}}
\end{figure}

Previous {\sl Chandra} observations of \object{G1.9+0.3} revealed
thermal emission lines from ejecta in the radio-bright northern rim
(Paper IV), 
but poor photon statistics precluded a search
for spatial variations in composition. We searched for
variations in the line emission within the 2011 May {\sl Chandra} data
cube.
First, this cube was denoised using the spectro-spatial method of
\citet{krishnamurthy10}.  Smoothed images at low, medium, and high
energies were extracted from a more heavily denoised data cube
containing all 2011 observations, and
combined into a color image (Figure \ref{2xim}). Spectral
variations are apparent, with the bright E and W lobes
noticeably harder than the much fainter S and N rims, confirming
azimuthal variations in the synchrotron spectra
(Paper III).

In our search for spectral lines within this cube, we used a
continuous wavelet transform method to spatially localize ejecta
emission.
We use the package {\tt MassSpecWavelet} developed by Du et al.~(2006)
for mass-spectrometry data. 
We concentrated on Si, S, and Fe, the most abundant ejecta elements whose
lines are prominent in young Type Ia SNRs, and which 
are also present in the large radio-bright N region 
of Paper IV. (We
do not confirm the presence of Ca and 
Sc there, and a very weak Ar line is not 
statistically significant.) For the Fe
K$\alpha$ line, we considered all possible emission line candidates
identified by {\tt MassSpecWavelet} in a relatively broad (6.12--6.76
keV) energy range near 6.4 keV in order to allow for potentially large
Doppler shifts.  For Si and S K$\alpha$ lines, our chosen energy
ranges were 1.65--1.97 keV and 2.15--2.53 keV, respectively.
In order to reduce the effects of noise, maps of estimated line
strengths were examined for spatially contiguous regions of potential
line emission. (In synchrotron-dominated SNRs where thermal and 
nonthermal emission is poorly correlated, this procedure could
result in regions not obviously corresponding to 
spatial structures visible in broad-band images.)
For these regions, we then extracted and combined X-ray
spectra from all 2011 observations. 
Here we discuss only selected regions with unambiguous  
line emission present, likely having highest density and shocked 
ejecta mass.

\begin{figure}
%\epsscale{1.0}
\epsscale{1.2}
\plotone{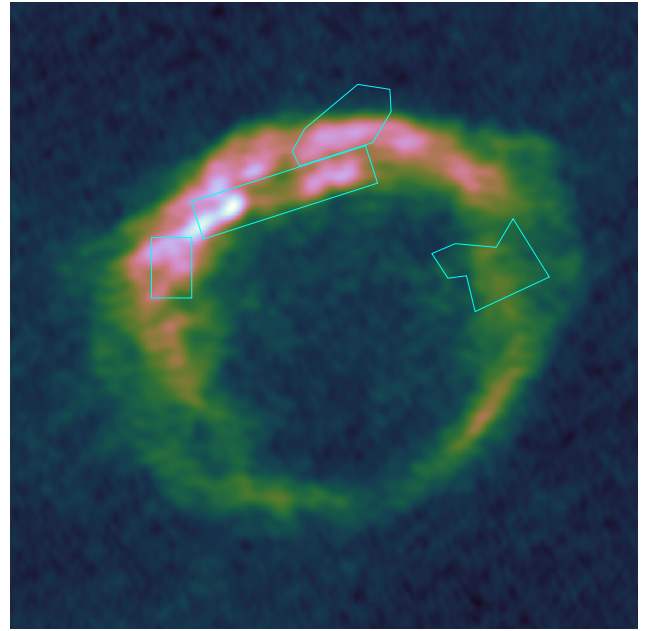}
%\plotone{g1p9radiodec2010.ps}
\caption{1.4 GHz VLA radio image from 2010 December (Green et al.~2013, in preparation), with regions containing spectral lines overlaid (top to bottom at 
left: outer N rim, inner N rim, NE region; at right, inner W rim).
Resolution $2.3'' \times 1.4''$. 
\label{radioimage}}
\end{figure}

\begin{figure*}
\epsscale{1.0} 
%\epsscale{1.18} 
\plotone{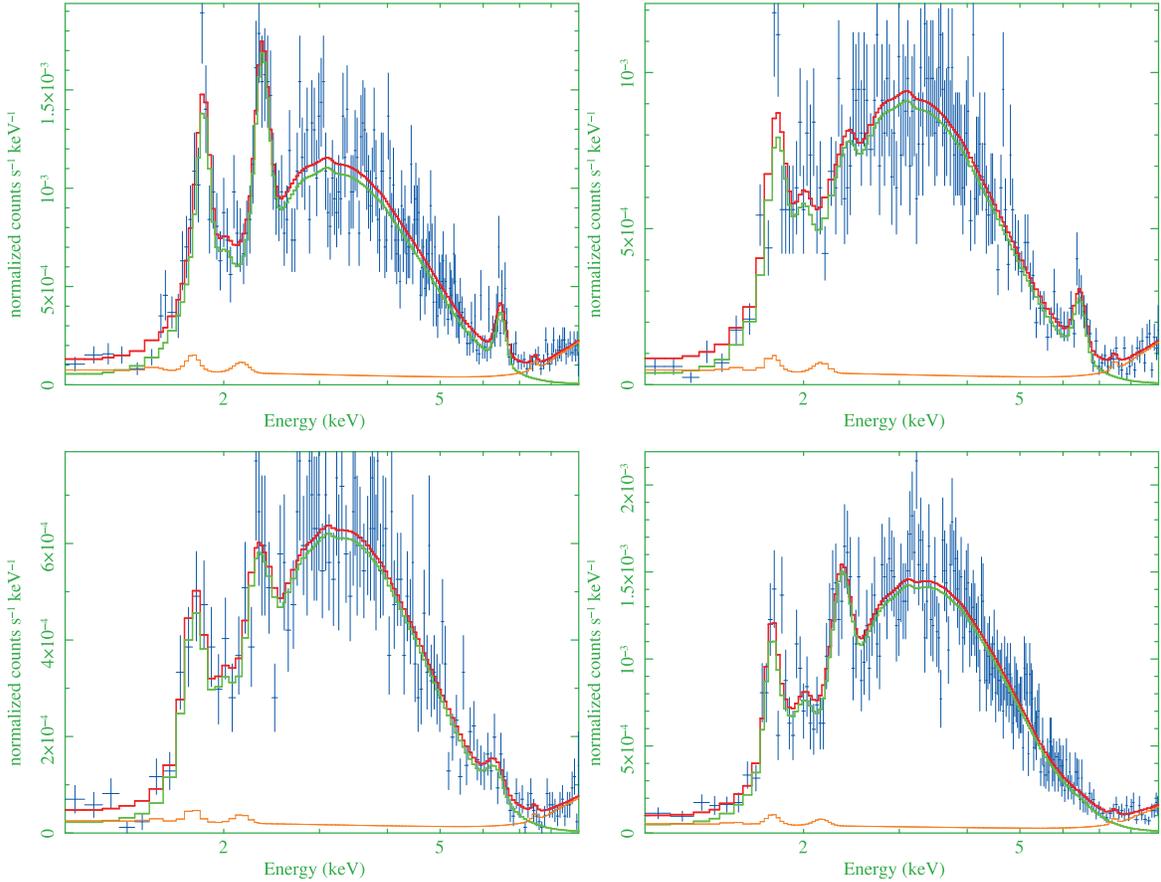}
%\plotone{spectra4.eps}
\caption{X-ray spectra. Top Left: Inner N rim.  Strong lines of Si, S, and Fe 
are present. Top Right: Outer N rim. Fe is present, but Si is weak and 
S is missing. Bottom Left: The NE region (without a soft X-ray 
point source visible in Figure 1). Si, S, and Fe are weaker than 
in inner N rim. Bottom Right: Inner W rim. Si and S are 
present, but Fe is missing.  
The orange and green lines are
the background and source models; the red line is the total.
\label{spec4}}
\end{figure*}

We identified 4 regions where either strong Fe
K$\alpha$ line and/or both Si and S K$\alpha$ lines are present (see
Figures \ref{2xim} and \ref{radioimage}).  Three regions mostly
overlap with the northern rim prominent in radio: inner
and outer northern rims, and a smaller northeastern region. This
confirms that most ejecta emission is associated with the radio-bright
N rim. The fourth region in the west overlaps with the innermost X-ray
and radio emission in this part of the remnant, demonstrating
that ejecta emission is present elsewhere than just the northern
rim. X-ray spectra are shown in Figure \ref{spec4}, with simple fits
consisting of absorbed power-law continua and Gaussian lines (see
Table \ref{lineflux} for fitted parameter values).
All lines are broad  
(FWHM $\sim 15,000$ km s$^{-1}$;
we assumed the same Doppler width for the IMEs and Fe).
Variations in X-ray spectra are
apparent: Fe K$\alpha$ is prominent in the northern rim
regions, but not in the western rim where only Si and S lines are
present. The outer northern rim exhibits strong Fe emission, but Si is
weak and S is missing. Both Fe and the IMEs (Si + S) are present in
the two remaining regions in the north, and are particularly strong in
the inner northern rim. 

\begin{deluxetable*}{lccccccc}
%\rotate
\tablecolumns{8}
\tablewidth{0pc}
%\tabletypesize{\scriptsize}
\tablecaption{Spectral Fits}

\tablehead{
\colhead{Region}  &$N_H$ &$\Gamma$\tablenotemark{a} &$F_{1-9\ {\rm keV}}$\tablenotemark{b} &FWHM &Silicon &Sulfur & Iron  \\
\colhead{} &($10^{22}$ cm$^{-2}$) & &($10^{-14}$ ergs cm$^{-2}$ s$^{-1}$) & ($10^4$ km s$^{-1}$) &\multicolumn{3}{c}{Line energies (keV) and strengths ($10^{-7}$ ph cm$^{-2}$ s$^{-1}$)}}

\startdata
\\
Outer north rim & 8.1 & 2.95 & 5.20 & 1.5 &1.756 & \nodata &6.452  \\
& (7.2, 9.0) & (2.64, 3.27) & (4.88, 5.54) 
&(0.75, 2.5) &(1.711, 1.792) & \nodata &(6.389, 6.518)  \\
& & & & &0.72 & \nodata &4.4  \\
& & & & &(0.32, 1.17) & \nodata &(2.8, 6.3)  \\

Inner north rim & 8.3 & 2.94 & 6.28 & 1.2 &1.822 &2.349 &6.475  \\
& (7.4, 9.2) & (2.66, 3.25) & (5.92, 6.65) 
&(0.81, 1.8) &(1.796, 1.849) &(2.327, 2.369) &(6.429, 6.523) \\
& & & & &1.7 &3.8 &5.5  \\
& & & & &(1.1, 2.3) &(2.8, 4.9) &(3.9, 7.3) \\

Inner NE rim & 9.0 & 2.96 & 3.79 & 2.1 &1.72 &2.27 &6.33  \\
& (7.7, 14) & (2.55, 4.27) & (3.14, 4.11) 
&(0.57, 6.2) &(1.56, 1.78) &(2.06, 2.34) &(5.98, 6.50) \\
& & & & & 0.65 & 1.1 & 1.6  \\
& & & & &(0.23, 1.7) &(0.38, 3.0) &(0.55, 4.9) \\

Inner west rim & 8.5 & 2.68 & 9.55 & 1.6 &1.738 &2.331 & \nodata \\
& (7.8, 9.4)& (2.46, 2.91) & (9.15, 9.98) 
&(0.56, 2.7) &(1.711, 1.762) &(2.294, 2.369) & \nodata \\
& & & & &1.4 &2.8 & \nodata \\
& & & & &(0.84, 1.9) &(1.7, 4.0) & \nodata

\enddata

\tablecomments{\ Line energies (strengths) are in rows 1, 5, 9, and 13 (3, 7, 
11, and 15), 
with 90\% confidence limits listed in adjacent rows.}
\tablenotetext{a}{Power-law photon index}
\tablenotetext{b}{Absorbed continuum flux in the 1--9 keV energy range}
\label{lineflux}
\end{deluxetable*}

Based on line identifications from the Gaussian fits, we fit {\tt
vpshock} models with variable abundances as given in Table \ref{lineflux}.  
Here we focus on the inner northern rim which has the richest line emission.
An absorbed {\tt vpshock} fit to its spectrum gives, with 90\% confidence
intervals, a hydrogen column density $N_H = 7.0 (6.3, 7.8) \times
10^{22}$ cm$^{-2}$, plasma temperature $kT$ of $3.2 (2.6, 3.9)$ keV,
ionization age $\tau = 2.4 (1.1, 4.0) \times 10^9$ cm$^{-3}$ s, and
the following abundances relative to solar values of \citet{grsa98}:
$3.0 (1.5, 5.1)$ for Si, $4.1 (2.8, 5.5)$ for S, and $7.4 (4.1, 12)$
for Fe. Lines are Doppler-broadened (FWHM $= 15000 (9200, 21000)$ km
s$^{-1}$) and blueshifted with a velocity of $2700 (600, 4900)$ km
s$^{-1}$.
Ejecta are blueshifted at a high confidence level of 98.0\%. 
In this fit, we used Markov chain Monte Carlo (MCMC) methods
(with uniform priors) instead of relying on C-statistics. Unlike the
latter, estimated parameter values and their errors obtained with MCMC
methods do not suffer systematic biases, and are more reliable. In
particular, comparison with the FWHM value of 12,000 (8100, 18,000) km
s$^{-1}$ for the inner N rim from Table \ref{lineflux} suggests that
some (or all) line widths listed there slightly underestimate true
line widths.

The NE rim spectrum is similar to the inner N rim 
spectrum, with best-fit values of 
$N_H = 7.3 (6.5, 9.1) \times 10^{22}$ cm$^{-2}$, 
$kT = 3.5 (2.2, 4.5)$ keV, $\tau = 1.0 (0, 4.8) \times 10^9$ cm$^{-3}$ s, 
Si = S $= 2.6 (1.2, 4.6)$, Fe $= 2.6 (0.8, 5.0)$. Lines might be redshifted 
by $4200 (-1200, 18000)$ km s$^{-1}$ and broad (FWHM $= 19000 (7600, 43000)$ 
km s$^{-1}$). While the fitted temperatures and ionization ages for the 
N and NE rims are similar, the relative velocity inferred 
from the fits is 6900 km $^{-1}$, and such high velocity is
supported by the large line broadenings required for both 
regions. We can thereby infer that the systematic line 
shifts for various regions seen in Table \ref{lineflux} are mostly caused 
by bulk motions, though ionization effects may still play a 
role.

In the {\tt vpshock} fit to the inner N rim spectrum, the X-ray
continuum is produced by free-free emission on protons and $\alpha$
particles, and we set the abundances of Ar and Ca equal to S.
Strongly oversolar abundances of Si, S, and Fe confirm that emission
lines are produced by SN ejecta. Mass emission measures (EMs) are $n_e
M_{Si(S)} =4.7 (4.5) \times 10^{-4}$ cm$^{-3}$ $M_\odot$ for the IMEs,
and $n_e
M_{Fe} =2.1 \times 10^{-3}$ cm$^{-3}$ $M_\odot$ for Fe.
The total mass EM is equal to 0.22 cm$^{-3}$ $M_\odot$,
and with $n_e \approx 2 \tau / t_{SNR} = 1.5$ cm$^{-3}$, the inferred
shocked mass is 0.14 $M_\odot$, several times larger than the
spatially-integrated shocked ejecta mass of 0.033 $M_\odot$ (derived from 
the overall remnant's dynamics, and assuming a standard Type Ia explosion; 
Paper V, 
see also Section \ref{ironejecta}). This
discrepancy suggests that the assumption that the continuum is thermal is
incorrect; given a nonthermal continuum,
it is likely that pure heavy-element ejecta, without any H or He,
produce the lines seen in Figure \ref{spec4}. This makes Si, S, and Fe
EMs somewhat uncertain, particularly for Fe if
plasma temperatures are significantly higher than 3.2 keV. As an
example, a fit to the inner N rim with a power-law model and a plane
shock model with an assumed temperature of 10 keV and without any H or
He gives EMs of 
$n_e M_{Si (S)} = 4.6 (2.6) \times 10^{-4}$ cm$^{-3}$ $M_\odot$
for the IMEs, and $2.4 \times 10^{-4}$ cm$^{-3}$ $M_\odot$ for 
Fe. A similar fit to the outer N rim spectrum gives $1.7 \times
10^{-4}$ cm$^{-3}$ $M_\odot$ for Si and $2.1 \times 10^{-4}$ cm$^{-3}$
$M_\odot$ for Fe.  In the W rim, $n_e M_{Si (S)} = 8.9 (2.6) \times
10^{-4}$ cm$^{-3}$ $M_\odot$, with $\tau = 0.14 (0, 1.4) \times 10^9$
cm$^{-3}$ s and velocity of $-1200 (-3200, 3100)$ km s$^{-1}$. 
Since the upper bound of $1.4 \times 10^9$ cm$^{-3}$ s on $\tau$ is less 
than our best estimate of ionization age in the inner N rim, ejecta 
densities 
likely decrease from N to W, and then continue decreasing South where ejecta 
emission becomes much fainter.

\section{High-velocity Fe-rich Ejecta} \label{ironejecta}

The presence of shocked heavy-element ejecta in \object{G1.9+0.3} is surprising, since all
ejecta that have been shocked so far were expelled by the SN with
velocities exceeding $\sim 18,000$ km s$^{-1}$ 
(Paper V).
In particular, Fe (or $^{56}$Ni) is not commonly found at such high
velocities in SN spectra, except for unusually energetic Type Ia SNe
and rare core-collapse (CC) hypernovae. In \object{G1.9+0.3}, Fe is prominent only
in the northern rim, not in the western rim where only Si and S lines
have been detected, implying a strongly asymmetric SN explosion. 
There is no evidence for any jet-like features expected in
jet-triggered energetic CC SNe, and other considerations (Paper I)
also favor a Type Ia origin, so in the following discussion we focus
on Type Ia explosions.

Studies of high-velocity Fe-group elements in Type Ia SN spectra are
rare since they require frequent acquisition of optical and UV spectra
soon after the SN explosion. The most detailed study so far is of a
normal albeit bright Type Ia SN, SN~2010jn \citep{hachinger13}, with a
broad light curve indicating substantial amounts of $^{56}$Ni. In
their preferred SN model, \cite{hachinger13} found good agreement with
observations only when they empirically adjusted abundances in the
delayed detonation model WS15DD3 of \citet{iwamoto99}. Fe-group
elements are present even at velocities exceeding 30,000 km s$^{-1}$;
at 20,000 km s$^{-1}$, the mass fraction of $^{56}$Ni and decay
products is $\sim 5$\%, with an additional $\sim 2$\%\ of directly
synthesized Fe present. No published Type Ia model has enough 
high-velocity Fe-group elements to account for their presence in
SN~2010jn.

We can use the results of Hachinger et al.~(2013) to estimate the mass
and EM of shocked ejecta in \object{G1.9+0.3}.  The density structure
in the WS15DD3 model can be approximated well by an exponential ejecta
model of \citet{dwarkadas98} with the explosion energy $E_{51}$ (in
units of $10^{51}$ ergs) of $1.43$ and the ejecta mass equal to the
Chandrasekhar mass.  The ejecta mass $M_{ej} (v)$ for free-expansion
velocities larger than $v$ is equal to $M_{ej}/M_{ej}^{tot} = 0.5
[(v/v_e)^2 + 2 v/v_e + 2] \exp ( -v/v_e )$, where $M_{ej}^{tot}$ is
the total ejecta mass and $v_e$ is the exponential velocity scale
($\rho_{ej} \propto \exp (-v/v_e)/t^3$). With $E_{51}=1.43$ and
$M_{ej}^{tot} = 1.38 M_\odot$, $v_e=2940$ km s$^{-1}$ instead of 
$v_e = 2440$ km s$^{-1}$ (for assumed $E_{51}=1$) in Paper V.
By repeating their
analysis for this more energetic SN model, we arrive at $v/v_e = 6.4$,
$v=19,000$ km s$^{-1}$, and $\rho_{ej} = 0.12$ amu cm$^{-3}$ for
ejecta entering the reverse shock in \object{G1.9+0.3} at the present time. The
total shocked ejecta mass is equal to 0.064 $M_\odot$, nearly twice as large as 
0.033 $M_\odot$ in the model of Paper V with $v/v_e = 7.3$.
We estimated EMs at 
$EM_{Si} \sim 6 \times 10^{-4}$ cm$^{-3}$ $M_\odot$, 
$EM_{S} \sim 1 \times 10^{-4}$ cm$^{-3}$
$M_\odot$, and $EM_{^{56}Ni+Fe} \sim 3 \times 10^{-4}$ cm$^{-3}$ $M_\odot$, using 
abundances of 15\%, 3\%, and 7\%  \citep[by mass;][]{hachinger13}, respectively.
These
spatially-integrated EMs are comparable to the EMs
we find in the inner northern rim (for a temperature of 10
keV), perhaps suggesting that ejecta there are 
several times denser than in the WS15DD3 model.

\section{Discussion}

Most Type Ia SN models lack the high-velocity Fe-group elements seen
in SN~2010jn and \object{G1.9+0.3}. \citet{meakin09} report the presence of
substantial ($>10$\%\ by mass) amounts of $^{56}$Ni up to velocities
of 18,500 km s$^{-1}$ in their gravitationally confined detonation
models. High-velocity (up to 18,000 km s$^{-1}$) $^{56}$Ni is present
in only one of a dozen 3D delayed-detonation models of
\citet{seitenzahl13}. In this model (N3), deflagrations were initiated
in only 3 central locations within the white dwarf (WD), producing a
strongly asymmetric and energetic explosion that synthesized nearly
$0.7 M_\odot$ of $^{56}$Ni.  All explosions with a small number of
ignition kernels resulted in energetic explosions, but fast-moving
($v \ga 12,000$ km s$^{-1}$) $^{56}$Ni is present only in models N3
and N5. In both models, part of a deflagration plume had already
risen to the surface of the WD when the first delayed detonation
occurred. The outermost deflagration ashes present within this plume
were then accelerated to high velocities by the SN shock wave.
\object{G1.9+0.3} might have been produced by such an explosion.
The strongly asymmetric distribution of Fe in \object{G1.9+0.3} supports this
hypothesis, since it is in qualitative agreement with the asymmetric
distribution of $^{56}$Ni in the N3 model. (The absence of strong azimuthal
variations in expansion rate 
(Paper V) 
disfavors an external
density gradient as an alternative explanation.) 

The ejecta emission in \object{G1.9+0.3} appears stronger than expected in
the heuristic SN~2010jn model of \citet{hachinger13}. One weakness of
that model is the assumption of spherical symmetry. The SN explosion could 
have been strongly asymmetric, as in the N3 model or in the
gravitationally-confined detonation models of \citet{meakin09} where
the density
profiles are much steeper in the direction of the
off-center detonation site than on the other side of the WD, leading
to an order of magnitude variation in density of high-velocity ejecta.
This could explain the lack of perfect agreement between
predictions based on the heuristic model of \cite{hachinger13} for
SN~2010jn and {\it Chandra} observations of \object{G1.9+0.3}.  The 
heavy-element ejecta in the northern
and western rims of \object{G1.9+0.3} were probably several times denser than in
the spherically-symmetric WS15DD3 model, perhaps because of their
location opposite to the off-center detonation site.

The coincidence of the radio maximum with a region of substantial Fe
suggests the possibility of a contribution to the radio flux from
positrons from the decay of $^{56}$Co injected into the
shock-acceleration process with initial energies of $~\sim$ 1 MeV.
20\% of $^{56}$Co decays produce a positron \citep{be06}; the $\sim 10^{-3}\
M_\odot$ of initial $^{56}$Ni we infer from the inner N rim will
eventually produce a total of $4 \times 10^{51}$ positrons.  However,
those injected at early times will lose energy in the dense ejecta due
to ionization losses;
once thermalized, they will rapidly annihilate.  \citet{martin10}
estimate the slowing-down time for positrons emitted at time $t$ as
$t_{\rm sd} \sim 10^{-3} E_{51}^{3/2}/M_{\rm ej}^{5/2} t^3$ yr, with
$M_{\rm ej}$ in $M_\odot$.  Then positrons emitted earlier than
$t_{\rm esc}$ given by $t_{\rm sd} = t$ will annihilate.  
For $E_{51} = 1$ and $M_{\rm ej} = 1.4\ M_\odot$, $t_{\rm esc}
\sim 50$ yr.  Given the 111 d mean life of $^{56}$Co, an
insignificant fraction of positrons is expected to survive to late
times, unless they are accelerated first.  The observed radio
flux from the inner N rim is about 150 mJy at 1.4 GHz; assuming a spectral
index of 0.62 (Paper III) and a power-law distribution of electrons
down to a minimum Lorentz factor of 10, standard synchrotron
formulae give
$$N_e = 2.7 \times 10^{50} (B/100\ \mu{\rm G})^{-1.62}\ {\rm electrons}$$
where nonlinear shock amplification would be required to reach
magnetic field strengths of order 100 $\mu$G.  For this value, all
positrons emitted after about 300 days would need to survive and be
accelerated.  This seems exceptionally unlikely.  
We conclude that decay positrons
cannot account for the radio maximum in the N rim; further radio
observations under way may cast light on this issue.

\object{G1.9+0.3} offers us a unique view
of the outermost, strongly asymmetric Fe-rich ejecta of a likely Type
Ia SNR. Their spatial distribution should allow for
distinguishing among various Type Ia explosion models.  Energetic
delayed-detonation models are promising candidates as they may account
for both the observed asymmetry and the presence of high-velocity
Fe. The delayed-detonation N3 model of \citet{seitenzahl13} produces
asymmetric ejecta rich in Fe-group elements, although not at high
enough free-expansion velocities. 
High-velocity radioactive $^{56}$Ni affects early-time Type Ia SN
light curves \citep{piro12,pironakar13,pironakar12}. Its highly
asymmetric spatial distribution, as observed in \object{G1.9+0.3} and
obtained in recent 3D Type Ia simulations such as the N3 model, should
be taken into account when interpreting light curves and spectra of
Type Ia SNe. Future 3D hydrodynamical simulations of asymmetric Type
Ia explosions, including additional delayed-detonation models,
followed by modeling of their collision with the ambient ISM, would be
most helpful for understanding the most recent known SN explosion in
our Galaxy. Such strongly asymmetric explosions are expected to
produce azimuthal variations in expansion rate that can be measured
with future {\sl Chandra} and VLA observations. In a broader context,
studies of this and other young Type Ia SNRs that show large-scale
asymmetries in their Fe-rich ejecta distribution 
\citep[e.g., SN 1006 and Kepler's SNR,][]{yamaguchi08,uchida13,burkey13} 
are invaluable for
understanding Type Ia explosions and their mysterious progenitors.

\acknowledgments

This work was supported by NASA through {\sl Chandra} General Observer
Program grants SAO G01-12098A and B.


\begin{thebibliography}{}

\bibitem[Arnaud(1996)]{arnaud96}
Arnaud, K. A. 1996, in Astronomical Data Analysis and Systems V, 
eds. G.Jacoby \& J.Barnes, ASP Conf.~Series, v.101, 17

\bibitem[Badenes et al.(2006)]{badenes06}
Badenes, C., Borkowski, K. J., Hughes, J. P., et al.
2006, ApJ, 645, 1373

\bibitem[B\'{e} et al.(2006)]{be06}
B\'{e}, M.-M., Chist\'{e}, V., Dulieu, C., et al. 2006, Table of 
Radionuclides, Monographie BIMP-5, Vol. 3 (S\`{e}vres: 
Bureau International des Poids et Mesures), 11

\bibitem[Borkowski et al.(2010, Paper IV)]{borkowski10}
Borkowski, K. J., Reynolds, S. P., Green, D. A., et al.
2010, ApJ, 724, 161 (Paper IV)

\bibitem[Borkowski et al.(2013)]{borkowski13}
Borkowski, K. J., Reynolds, S. P., Green, D. A., et al.
2013, Proc. of the 13th HEAD meeting of the AAS

\bibitem[Burkey et al.(2013)]{burkey13}
Burkey, M. T., Reynolds, S. P., Borkowski, K. J., et al.
2013, ApJ, 764. 63 

\bibitem[Carlton et al.(2011, Paper V)]{carlton11}
Carlton, A. K., Borkowski, K. J., Reynolds, S. P., et al. 
2011, ApJ, 2011, 737, 22 (Paper V)

\bibitem[Clark et al.(1991)]{clark91}
Clark, R. E. H.,  Abdallah, J., Jr., \& Mann, J. B. 1991, ApJ, 381, 597

\bibitem[Du et al.(2006)]{du06}
Du, P., Kibbe, W. A., \& Lin, S. M.
2006, Bioinformatics, 22, 2059

\bibitem[Dwarkadas \& Chevalier(1998)]{dwarkadas98}
Dwarkadas, V. V., \& Chevalier, R. A. 1998, ApJ, 497, 807

\bibitem[Green et al.(2008, Paper II)]{green08}
Green, D. A., Reynolds, S. P., Borkowski, K. J., 
et al.
2008, MNRAS, 387, L54 (Paper II)

\bibitem[Grevesse \& Sauval(1998)]{grsa98}
Grevesse, N., \& Sauval, A. J.
1998, Space Sci.~Rev., 85, 161

\bibitem[Hachinger et al.(2013)]{hachinger13}
Hachinger, S., Mazzali, P. A., Sullivan, M., et al.
2013, MNRAS, 429, 2228

\bibitem[Iwamoto et al.(1999)]{iwamoto99}
Iwamoto, K., Brachwitz, F., Nomoto, K., 
et al.
1999, ApJS, 125, 439

\bibitem[Krishnamurthy et al.(2010)]{krishnamurthy10}
Krishnamurthy, K., Raginsky, M., \& Willett, R. 2010, SIAM J. Imaging Sci., 3, 619

\bibitem[Mann(1983)]{mann83}
Mann, J. B. 1983, At.~Dat.~Nuc.~Data Tables, 29, 407

\bibitem[Martin et al.(2010)]{martin10}
Martin, P., Vink, J., Jiraskova, S., 
et al. 2010, A\&A, 519:A100

\bibitem[Meakin et al.(2009)]{meakin09}
Meakin, C. A., Seitenzahl, I., Townsley, D., et al. 2009, ApJ, 693, 1188

\bibitem[Piro(2012)]{piro12}
Piro, A. L.
2012, ApJ, 759, 83

\bibitem[Piro \& Nakar(2013)]{pironakar13}
Piro, A. L., \& Nakar, E. 2013, ApJ, 769, 67

\bibitem[Piro \& Nakar(2012)]{pironakar12}
Piro, A. L., \& Nakar, E. 2012, preprint, arXiv:1211.6438

\bibitem[Palmeri et al.(2008)]{palmeri08}
Palmeri, P., Quinet, P., Mendoza, C., et al. 2008, ApJS, 177, 408

\bibitem[Reynolds et al.(2008, Paper I)]{reynolds08b}
Reynolds, S. P., Borkowski, K. J., Green, D. A., 
et al.
2008, ApJ, 680, L41 (Paper I)

\bibitem[Reynolds et al.(2009, Paper III)]{reynolds09}
Reynolds, S. P., Borkowski, K. J., Green, D. A., 
et al.
2009, ApJ, 695, L149 (Paper III)

\bibitem[Seitenzahl et al.(2013)]{seitenzahl13}
Seitenzahl, I. R., Ciaraldi-Schoolmann, F., R\"{o}pke, F. K., et al. 
2013, MNRAS, 429, 1156

\bibitem[Uchida et al.(2013)]{uchida13}
Uchida, H., Yamaguchi, H., \& Koyama, K. 2013, ApJ, in press

\bibitem[Yamaguchi et al.(2008)]{yamaguchi08}
Yamaguchi, H., Koyama, K., Katsuda, S., et al. 2008, PASJ, 60, 141


\end{thebibliography}
\end{document}